# Exploring the Potential of Spiking Neural Networks in UWB NLOS Identification

Youdong Zhang†, *Graduate Student Member, IEEE*, Xu He†, *Graduate Student Member, IEEE*, Xiaolin Meng*

***Abstract*—Ultra-Wideband (UWB) Non-Line-of-Sight (NLOS) identification is critical for accurate localization, yet prevailing deep learning solutions impose computational demands incompatible with low-cost edge devices. This letter explores the potential of Spiking Neural Networks (SNNs) for this task and develops a fully unsupervised SNN solution. Against supervised deep learning baselines, our approach achieves comparable accuracy (~82%) with only 4.7K synaptic operations per inference, representing a three-order-of-magnitude improvement in efficiency. These results establish that SNNs can serve as a viable neuromorphic alternative for edge UWB applications.**

## I. Introduction

UWB plays a vital role in applications spanning intelligent mobility, precision agriculture, smart homes, etc. [1]. In practice, detecting UWB NLOS channels is essential for filtering outliers and enabling precise compensation [2]. A large body of work has already embraced learning-based approaches, confirming the value of deep learning for UWB NLOS identification. Recent studies have concentrated on algorithmic innovations to further improve estimation accuracy [3], [4]. However, these highly complex models incur heavy computational loads, making them ill-suited for the low-cost embedded devices typical of UWB systems. Consequently, some studies stress the need for practical, resource-efficient learning-based solutions when deploying UWB NLOS identification at the edge [2], [5].

SNNs, known as the third generation of neural networks, more faithfully reproduce the dynamics of real neurons than Artificial Neural Networks (ANNs) [6]. Although training SNNs is more challenging, their potential for dramatic gains in computational efficiency and ultra-low-power neuromorphic deployment is a major treasure [7]. To our knowledge, virtually no work has explored SNN-based solutions for UWB NLOS Identification. Only [8] and [9], which address SNN-based classification of Impulse Radio (IR)-UWB radar signals, far away from the standard UWB NLOS identification problem tackled here. This gap motivates our work.

To avoid the training difficulties with SNNs, we adopt the Liquid State Machine (LSM) [10], a recurrent SNN with fixed synaptic weights, to extract spiking representations of both UWB RF channel features and Channel Impulse Response (CIR) features after spike encoding. Specifically, we first engineer the UWB channel data into a form compatible with spike-based representation learning, and then develop a fully unsupervised SNN pipeline (Section II) for NLOS identification. Evaluated on the public eWINE benchmark [11], this work achieves performance comparable to that of supervised learning methods.

## II. Algorithm Design

The proposed fully unsupervised SNN method is illustrated in Figure 1. It consists of two main components: a feature engineering block and a liquid NLOS identification block.

### *A. Feature Engineering Block*

*RF Feature Processing and Spike Encoding*: The RF chip of the DW1000-based UWB module exposes RF features that can be used to analyze channel state (details in [1]). These features include statistical indicators of ranging observations and noise levels, first path metrics, channel-quality and energy metrics, forming a 10-dimensional feature vector. To make the data compatible with the subsequent SNN, this letter applies rate encoding [12] to convert the RF feature vector into spikes that serves as the input ($X_1$) to the RF liquid encoder.

*CIR Feature Processing and Spike Encoding*: Compared with the RF feature vector, the UWB CIR signal carries richer channel information. The CIR data format supplied by DW1000-based devices depends on the selected Pulse Repetition Frequency (PRF) [2]: at 64 MHz the CIR is delivered as a 1024-sample real sequence, whereas at 16 MHz it is a 992-sample complex sequence. This letter focuses on the real CIR sequence obtained with PRF = 64 MHz.

Earlier work [2] has shown that, for this configuration, the 50 samples starting from FP_IDX (register index of the first-arrival path peak) are the most discriminative for NLOS identification, a finding subsequently exploited in [4] and [10]. More recent studies [1] and the creators of the eWINE dataset [11] have argued that 120 samples starting from FP_IDX are beneficial for channel modeling. Therefore, we extract two candidate segments: 50-length and 120-length CIR vectors.

Following [5] , this letter also applies uniform padding with a spacing of 10 samples to both extracted segments. The padded vectors are then converted into spike trains via frequency encoding, yielding the input pattern (X2) for the CIR liquid encoder. The same padding and encoding procedures are applied regardless of segment length to ensure consistent experimental conditions.

This work was supported by the Natural Science Foundation of Jiangsu Province under Grant No. BK20243064. Xu He receives support from the Postgraduate Research & Practice Innovation Program of Jiangsu Province under Grant No. SJCX24_0067 and SEU Innovation Capability Enhancement Plan for Doctoral Students under Grant No. CXJH_SEU 24204. (*Corresponding author: Xiaolin Meng*)

Youdong Zhang and Xu He contributed equally to this work. Both are co-first authors. † indicates equal contribution.

The authors are with the School of Instrument Science and Engineering, the China-UK Research Centre on Intelligent Mobility, State Key Laboratory of Comprehensive PNT Network and Equipment Technology, Southeast University, Nanjing 210096, China (e-mail: xiaolin_meng@seu.edu.cn).

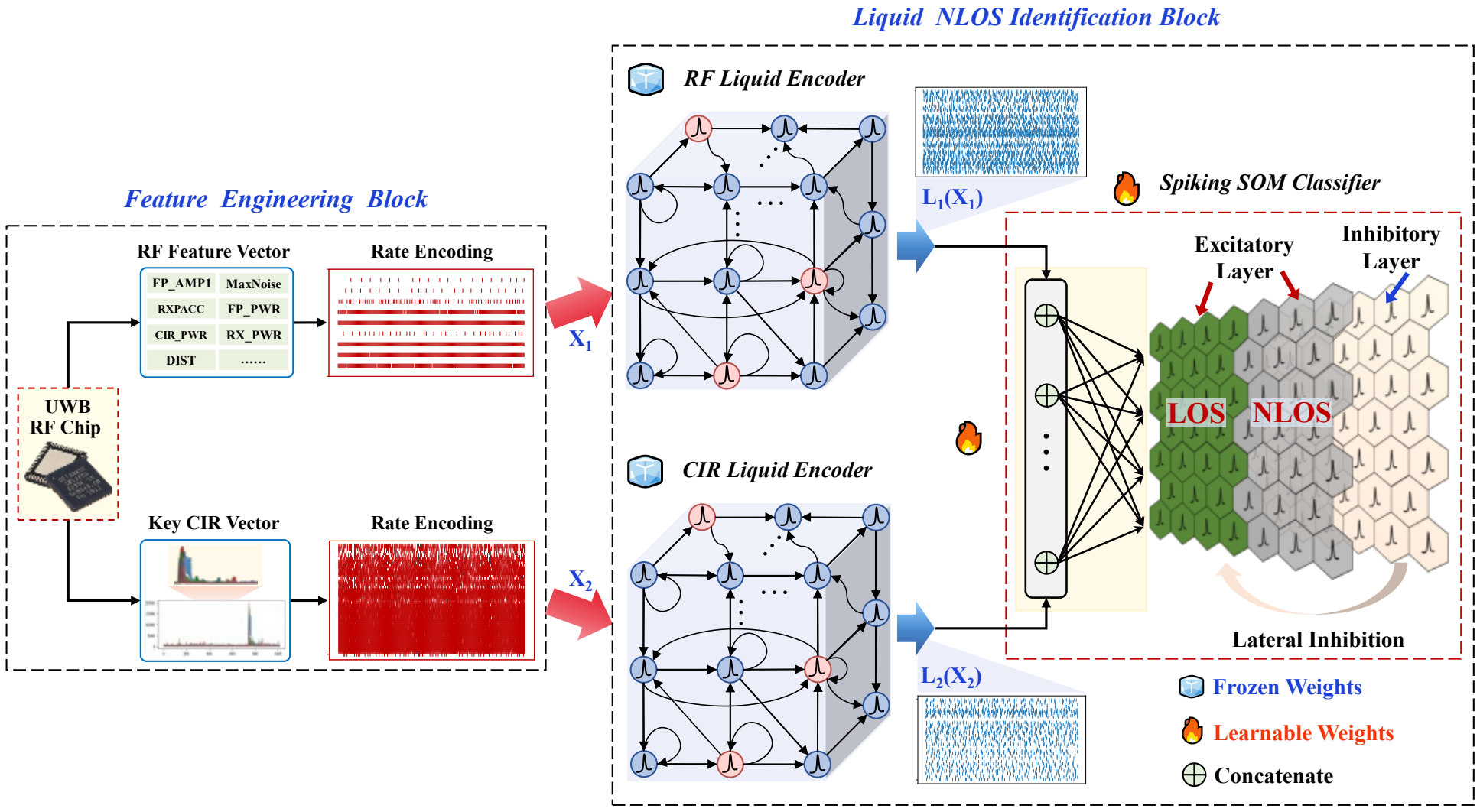

Fig. 1. The proposed method.

## B. Liquid NLOS Identification Block

*Liquid Encoders*: Many deep learning studies rely solely on CIR for UWB NLOS identification [3]. Yet, under complicated multipath environments, CIR alone often confuses NLOS with LOS signals. Some works advocate fusing RF features with CIR for more robust NLOS Identification[1], [2]. Accordingly, we feed the spike-encoded RF and CIR streams into two separate liquid encoders to extract their spiking patterns, denoted $L_1(X_1)$ and $L_2(X_2)$ in Figure 1.

As a spiking counterpart to recurrent networks, the liquids in LSM transform input features into high-dimensional spiking representations through a randomly recurrent layer of spiking neurons. Crucially, this encoding requires no training [10].

Both the RF and CIR liquid encoders are built with Leaky Integrate-and-Fire (LIF) neurons [7]. Each liquid encoder consists of excitatory and inhibitory neurons (color-coded in Figure 1). Synapses are randomly initialized, sparse, and include a fraction of self-connections.

Given the discrete time steps employed in our method, the dynamics of the LIF neuron are formulated as follows:

$$V[t] = V[t-1] - \frac{1}{\tau_m}(V[t-1] - V_{\text{rest}}) + WX[t] - S[t]\theta \quad (1)$$

where, $V[t]$ is the membrane potential, $\tau_m$ is the membrane time constant, $V_{\text{rest}}$ is the resting potential, $W$ are the synaptic weights, $X[t]$ is the input and $\theta$ is the threshold. Note that if a neuron is fired at time $t$, its membrane potential is reset at that moment. Therefore, the membrane voltage at time $t$ consists of the decay term $(V[t-1] - V_{\text{rest}})/\tau_m$, the input term $WX[t]$, and the reset term $S[t]\theta$. The spiking activation function $S[t]$ is defined in (2) as follows:

$$S[t] = \begin{cases} 1, & \text{if } V[t] \geq \theta \\ 0, & \text{otherwise} \end{cases} \quad (2)$$

The related LIF parameters follow [14]. Compared with many deep learning models that use convolutional operations or attention mechanisms [3], [4], the liquid encoders offer markedly lower computational cost and higher efficiency.

*Spiking SOM Classifier*: After the two liquid encoders produce their respective spike patterns, the resulting 1D tensors are concatenated into a joint spike pattern that serves as input to the spiking Self-Organizing Map (SOM) [13] classifier, a spike-based variant of the SOM. The spiking SOM is likewise implemented with LIF neurons.

Unlike traditional SOM, competition is decided by counting input spikes: the neuron receiving the largest number of spikes is declared the winner. The winner then strengthens the activation of its neighbors, which learn similar spike patterns under the same input, while simultaneously exerting lateral inhibition on non-neighboring neurons, reducing their probability of firing. Classification is performed by choosing the label whose associated neurons exhibit the highest average spike count per sample. Given the space constraints of the Letter format, details on the spiking SOM are in [13].

## III. Experiment and Analysis

*1) Experiment Setup.* We evaluate our method on the publicly available eWINE benchmark [11]. All LIF neurons are simulated with the open-source library BindsNET [14], to which the first author is a contributor. Experiments were run on a workstation with 32 CPU cores, 251 GB RAM, and an RTX 4090 GPU. The liquid encoders use fixed, untrained weights; only the spiking SOM is trained, using the unsupervised, biologically inspired Spike Timing Dependent Plasticity (STDP) rule [15]. The dataset is split 5:2 into training and test sets, with no overlap.

*2) Experimental Results and Ablation.* The performance results of our proposed method are shown in Table I, which includes two main strategies and eight ablation strategies.

- ❑ *Strategy1*: The 50-sample CIR segment is spike-encoded into $X_2$ and fed to the CIR liquid encoder to produce spike pattern $L_2(X_2)$. Following the pipeline in Figure 1, $L_2(X_2)$ is fused with the RF-branch spike pattern $L_1(X_1)$ and classified by the spiking SOM.
- ❑ *Strategy2*: Same as Strategy1, using a 120-sample CIR segment.
- ❑ *Strategy3*: Same as Strategy1, ablating CIR features.
- ❑ *Strategy4*: Same as Strategy3, ablating the RF liquid encoder.
- ❑ *Strategy5*: Same as Strategy1, ablating RF features.
- ❑ *Strategy6*: Same as Strategy5, ablating the CIR liquid encoder.
- ❑ *Strategy7*: Same as Strategy2, ablating RF features.
- ❑ *Strategy8*: Same as Strategy7, ablating the CIR liquid encoder.
- ❑ *Strategy9*: Same as Strategy1, ablating all liquid encoders.
- ❑ *Strategy10*: Same as Strategy2, ablating all liquid encoders.

TABLE I
EVALUATIONS OF ALL DESIGNED STRATEGIES ON THE EWINE BENCHMARK

| Strategies | Model Structures | | | | Metrics | | | |
|---|---|---|---|---|---|---|---|---|
| | Feature Engineering | RF Liquid Encoder | CIR Liquid Encoder | Spiking SOM Classifier | Accuracy | Precision | Recall | F1 Score |
| **Strategy1** | **RF+CIR(50)** | √ | √ | √ | **80.9%** | **82.3%** | **78.9%** | **80.6%** |
| **Strategy2** | **RF+CIR(120)** | √ | √ | √ | **81.8%** | **82.7%** | **80.8%** | **81.7%** |
| Strategy3 | RF | √ | × | √ | 77.0% | 81.8% | 69.1% | 74.9% |
| Strategy4 | RF | × | × | √ | 51.3% | 51.1% | 46.3% | 48.6% |
| Strategy5 | CIR(50) | × | √ | √ | 58.1% | 57.3% | 61.4% | 59.3% |
| Strategy6 | CIR(50) | × | × | √ | 50.1% | 49.8% | 53.7% | 51.7% |
| Strategy7 | CIR(120) | × | √ | √ | 60.3% | 58.9% | 66.9% | 62.6% |
| Strategy8 | CIR(120) | × | × | √ | 49.9% | 49.6% | 48.2% | 48.9% |
| Strategy9 | RF+CIR(50) | × | × | √ | 51.2% | 50.9% | 53.0% | 51.9% |
| Strategy10 | RF+CIR(120) | × | × | √ | 49.6% | 49.3% | 50.8% | 50.0% |

The evaluation metrics of all strategies are quantified in Table I. The results demonstrate that the proposed Strategy1 and Strategy2 outperform all other methods. The choice between 50 and 120 CIR samples has only a minor impact on accuracy, but the longer segment (Strategy2) is slightly stronger.

The liquid encoders are essential, as omitting them (Strategies 9 and 10) led to a ~30% reduction in all performance metrics compared to the Strategies 1 and 2. Moreover, removing the CIR liquid encoder from Strategy7 to form Strategy8 causes a drop of ~10% across all metrics. Similar trends are found when comparing Strategy 5 and Strategy 6, as well as Strategy 3 and Strategy 4.

The size of the liquid encoders is a key hyperparameter that determines the model's representation capability. To determine the optimal configuration, we evaluated the model's accuracy as the number of LIF neurons in the liquid encoders varied from 400 to 1,600. As shown in Figure 2, accuracy improved significantly as the number of neurons increased to 900. Beyond this threshold, the test accuracy no longer increases. However, the number of spikes continues to rise. Therefore, we selected 900 neurons to achieve the balance between encoding quality and the number of spikes.

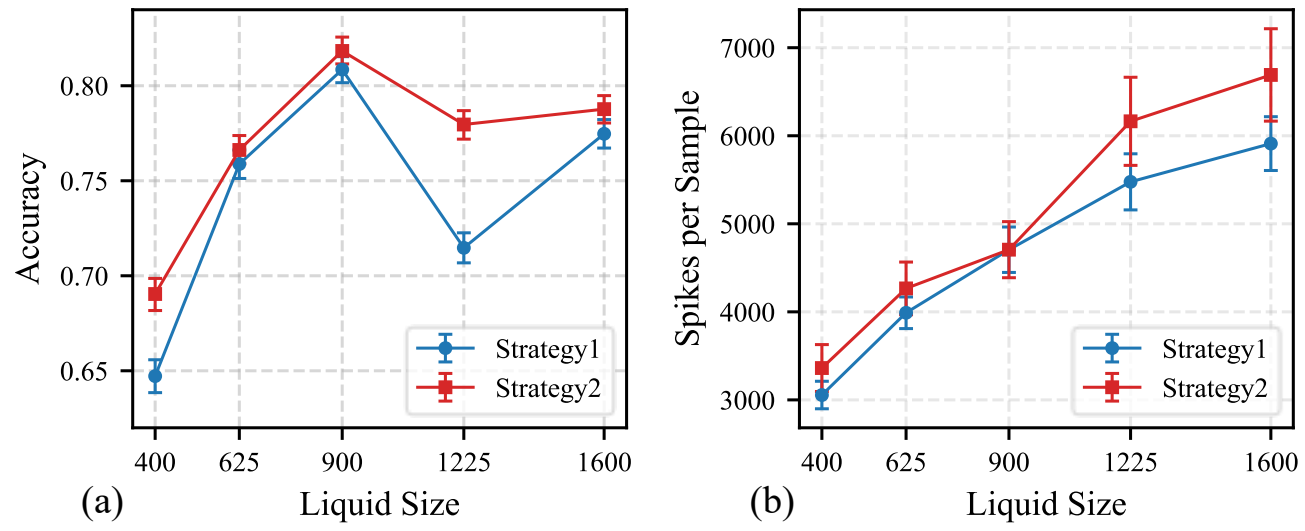

Fig. 2. The effects of liquid size. (a) Accuracy. (b) Spike counts.

*3) Experimental Analysis.* To evaluate computational efficiency, we conducted a quantitative analysis of the inference phase. Results indicate that the average inference latency is 0.2 ms. Furthermore, owing to the temporal sparsity of spikes, the network exhibits high sparsity during inference, requiring an average of only 4.7K Synaptic Operations (SOPs) per sample. Based on the energy profile of Intel's Loihi neuromorphic chip at 23.6 pJ/SOP [16], the energy consumption per inference is only 0.11 μJ.

To verify that the representations generated by the liquid encoders can effectively drive the competitive dynamics of the spiking SOM, we analyzed the spike statistics of the liquid encoders. As shown in Figure 3, the firing rate tends to stabilize over time steps. The average temporal sparsity reaches 97.9% and sparsity distribution approximates a normal distribution. This highly sparse and low-redundancy representation provides ideal input conditions for the lateral inhibition mechanism and winner-take-all (WTA) competition, enabling efficient and stable NLOS identification without complex joint parameter tuning.

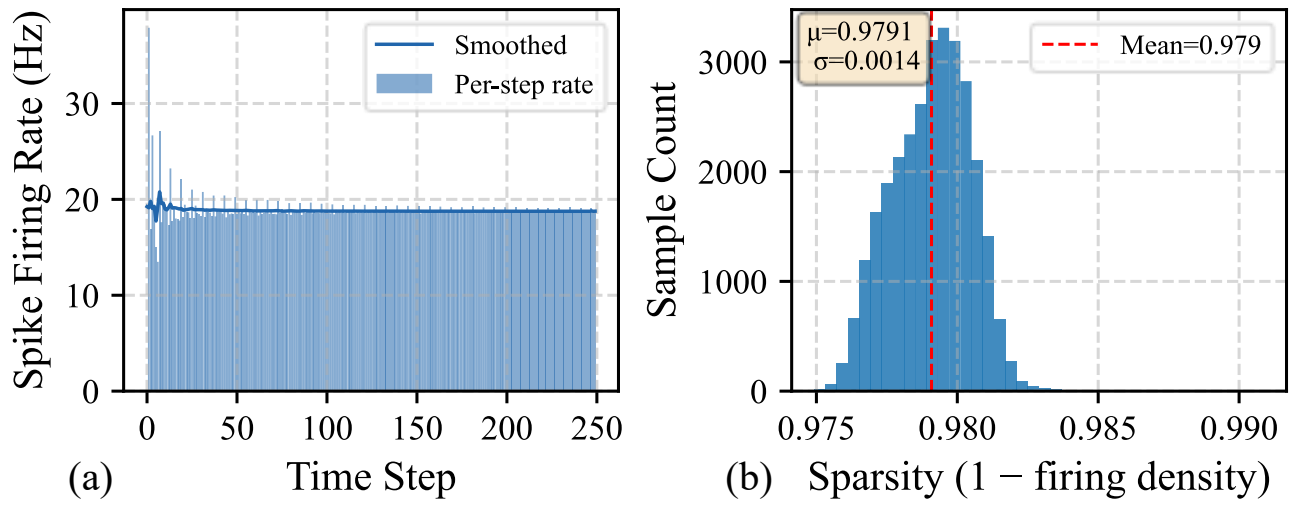

Fig. 3. (a) Spike firing rate. (b) Spike sparsity distribution.

The decision behavior reflects a favorable balance between energy efficiency and latency. Leveraging accumulated spike activity for classification, the network exhibits a notable early-decision capability. Most of the first spike occur within 50 time steps, and stable classification outputs are produced within the first 8 time steps (Figure 4). This early convergence pattern preserves NLOS identification accuracy while reducing inference response time and lowers the dynamic power consumption, further highlighting the deployment advantages of the proposed method.

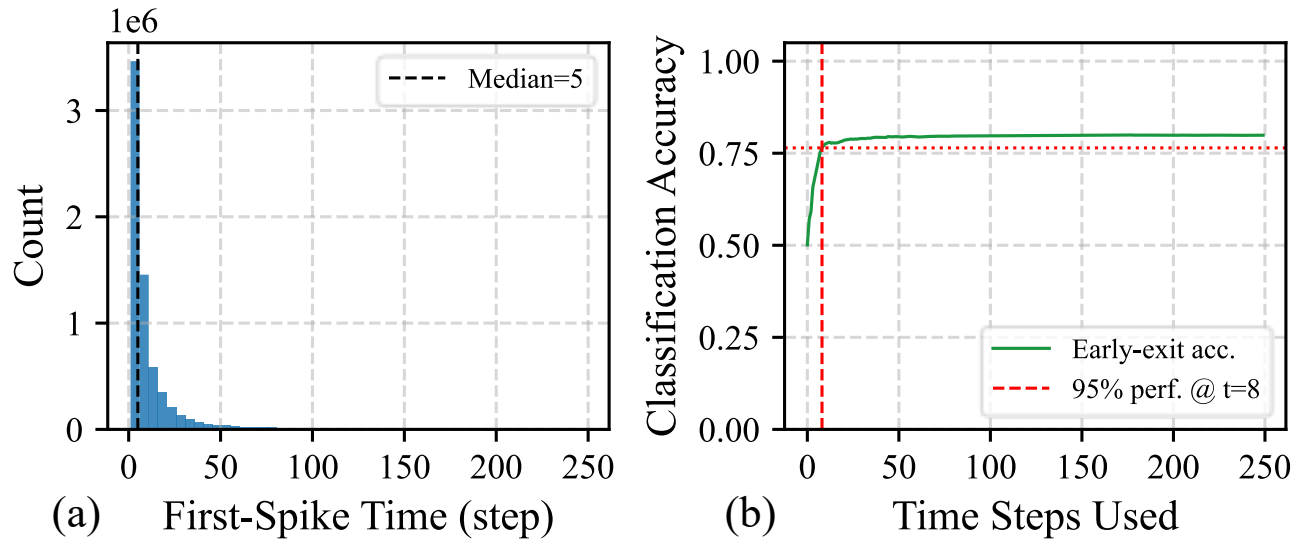

Fig. 4. (a) First-spike time distribution. (b) Early-exit decision curve.

To assess model uncertainty, we applied bootstrapping with 2,000 resamples to the test set. The resulting 95% confidence intervals for NLOS identification accuracy are reported in Table II. The narrow intervals indicate that the proposed method maintains highly stable performance across varying data distributions.

TABLE II
BOOTSTRAP 95% CONFIDENCE INTERVALS FOR ACCURACY

| Strategies | Liquid Size | | | | |
|---|---|---|---|---|---|
| | 400 | 625 | 900 | 1225 | 1600 |
| Strategy1 | [63.85%, 65.58%] | [75.12%, 76.62%] | [80.16%, 81.53%] | [70.68%, 72.26%] | [76.72%, 78.22%] |
| Strategy2 | [68.17%, 69.86%] | [75.90%, 77.38%] | [81.15%, 82.56%] | [77.19%, 78.69%] | [78.04%, 79.48%] |

*4) Comparative Analysis.* Additionally, this letter also includes several recent supervised deep learning baselines for comparison. For instance, Jiang et al. [17] reports multiple baseline results on the eWINE dataset, including classic multi-layer CNNs and CNN-LSTM models with different Long Short-Term Memory (LSTM) architectures. Deng et al. [18] presents WGCNN, a modified CNN strategy, for NLOS identification. Due to space constraints, the details of all baselines are omitted here; please refer to their original works [17], [18]. The results show that our fully unsupervised SNN scheme achieves accuracy close to that of traditional supervised deep learning methods. Although this fully unsupervised approach achieves accuracy only comparable to that of highly complex deep learning networks, it yields substantial dividends in energy efficiency. Consequently, this trade-off positions the SNN not as a direct accuracy competitor, but as a computationally efficient, biologically interpretable neuromorphic alternative for low-latency and low-power UWB edge sensing.

Furthermore, the proposed model adopts a two-layer architecture with less than 0.21M trainable parameters. To evaluate computational complexity, we use Floating-Point Operations (FLOPs) for deep learning models and Synaptic Operations (SOPs) for our model, the latter being more appropriate given the binary nature of SNNs. Owing to the inherently high temporal sparsity of the liquid encoders, the model executes on average only 4.7K SOPs per inference, which is three orders of magnitude lower than the FLOPs required by deep learning baselines (see Table III).

TABLE III
COMPARISON WITH DEEP LEARNING MODELS

| Models | Accuracy | Layers | Parameters | FLOPs |
|---|---|---|---|---|
| CNN [17] | 81.14% | 5 | ~0.64M | ~1.2M |
| CNN-LSTM [17] | 81.56% | 6 | ~0.66M | ~11.5M |
| CNN-Bi LSTM [17] | 78.93% | 6 | ~0.68M | ~21.7M |
| CNN-Stacked LSTM [17] | 82.14% | 6 | ~0.67M | ~21.7M |
| WGCNN [18] | 82.63% | 6 | ~0.50M | ~3.0M |
| **This Work** | **81.80%** | **2** | **<0.21M** | **4.7K*** |

*Reported as 4.7K SOPs.

## IV. DISCUSSION AND CONCLUSION

This letter explores the potential of SNNs for UWB NLOS identification, proposing a fully unsupervised approach that achieves ~82% accuracy at a computational cost of 4.7K synaptic operations, comparable to deep learning methods while substantially reducing computational complexity. The spiking SOM classifier, trained via STDP, offers greater simplicity and efficiency compared to end-to-end training of complex SNNs. Although ANN-to-SNN conversion (e.g., SpikingJelly [19]) represents a widely adopted strategy for transferring pretrained model weights, such approaches fundamentally map continuous ANN activations to spike firing rates, lacking the biologically meaningful synaptic plasticity dynamics inherent to genuine spiking neural computation. In contrast, the fully unsupervised SNN framework proposed in this paper offers biological interpretability.

This work introduces the neuromorphic computing paradigm into the wireless sensing community from the perspective of the intersection between computational neuroscience and communications, thereby offering value for advancing future low-power neuromorphic computing in IoT edge devices.

*Note.* Data and code for this work are available at https://github.com/BINUCOE/SpikingUWB.